\documentclass[10pt,twocolumn,amssymb, amsmath, aps, superscriptaddress, showpacs, footinbib, prb]{revtex4-1}
\usepackage{graphicx}

\def\be{\begin{equation}}
\def\ee{\end{equation}}
\def\bea{\begin{eqnarray}}
\def\eea{\end{eqnarray}}

\def\vec{\mathbf}
\def\mc{\mathcal}

\newcommand{\eff}{\text{eff}}
\newcommand{\AFM}{\text{AFM}}
\newcommand{\FM}{\text{FM}}

\newcommand{\Dv}{\mathbf D}

\begin{document}

\title{Bridging frustrated-spin-chain and spin-ladder physics:\\ quasi-one-dimensional magnetism of BiCu$_2$PO$_6$}

\author{Alexander A. Tsirlin}
\email{altsirlin@gmail.com}
\affiliation{Max-Planck-Institut f\"ur Chemische Physik fester Stoffe, 01187 Dresden, Germany}
\author{Ioannis Rousochatzakis}
\email{rousocha@pks.mpg.de}
\affiliation{Max-Planck-Institut f\"ur Physik komplexer Systeme, 01187 Dresden, Germany}
\author{Deepa Kasinathan}
\author{Oleg Janson}
\affiliation{Max-Planck-Institut f\"ur Chemische Physik fester Stoffe, 01187 Dresden, Germany}
\author{Ramesh Nath}
\affiliation{Max-Planck-Institut f\"ur Chemische Physik fester Stoffe, 01187 Dresden, Germany}
\affiliation{Indian Institute of Science Education and Research, Trivandrum-695016 Kerala, India}
\author{Franziska Weickert}
\affiliation{Max-Planck-Institut f\"ur Chemische Physik fester Stoffe, 01187 Dresden, Germany}
\affiliation{Dresden High Magnetic Field Laboratory, Forschungszentrum Dresden-Rossendorf, 01314 Dresden, Germany}
\author{Christoph Geibel}
\affiliation{Max-Planck-Institut f\"ur Chemische Physik fester Stoffe, 01187 Dresden, Germany}
\author{Andreas M. L\"auchli}
\email{aml@pks.mpg.de}
\affiliation{Max-Planck-Institut f\"ur Physik komplexer Systeme, 01187 Dresden, Germany}
\author{Helge Rosner}
%\email{Helge.Rosner@cpfs.mpg.de}
\affiliation{Max-Planck-Institut f\"ur Chemische Physik fester Stoffe, 01187 Dresden, Germany}
%\date{\today}

\begin{abstract}
We derive and investigate the microscopic model of the quantum magnet BiCu$_2$PO$_6$ using band structure calculations, magnetic susceptibility and high-field magnetization measurements, as well as Exact Diagonalization (ED) and Density-Matrix Renormalization Group (DMRG) techniques. The resulting quasi-one-dimensional spin model is a two-leg antiferromagnetic ladder with frustrating next-nearest-neighbor couplings along the legs. The individual couplings are estimated from band structure calculations and by fitting the magnetic susceptibility with theoretical predictions, obtained using full diagonalizations. The nearest-neighbor leg coupling $J_1$, the rung coupling $J_4$, and one of the next-nearest-neighbor couplings $J_2$ amount to $120-150$~K, while the second next-nearest-neighbor coupling is $J_2'\simeq J_2/2$. The spin ladders do not match the structural chains, and although the next-nearest-neighbor interactions $J_2$ and $J_2'$ have very similar superexchange pathways, they differ substantially in magnitude due to a tiny difference in the O--O distances and in the arrangement of non-magnetic PO$_4$ tetrahedra. An extensive ED study of the proposed model provides the low-energy excitation spectrum and shows that the system is in the strong {\em rung} coupling regime. The strong frustration by the next-nearest-neighbor couplings leads to a triplon branch with an incommensurate minimum. This is further corroborated by a strong-coupling expansion up to second order in the inter-rung coupling. Based on high-field magnetization measurements, we estimate the spin gap of $\Delta\simeq 32$~K and suggest the likely presence of antisymmetric Dzyaloshinskii-Moriya anisotropy and inter-ladder coupling $J_3$. We also provide a tentative description of the physics of BiCu$_2$PO$_6$ in magnetic field, in the light of the low-energy excitation spectra and numerical calculations based on ED and DMRG. In particular, we raise the possibility for a rich interplay between one- and two-component Luttinger liquid phases and a magnetization plateau at 1/2 of the saturation value.  
\end{abstract}

\maketitle

\section{Introduction}
One-dimensional (1D) spin systems are in the focus of the present-day research due to a range of unusual low-temperature properties governed by quantum effects. The primary 1D spin model is the uniform spin-$\frac12$ Heisenberg chain that has a peculiar gapless excitation spectrum.\cite{mueller1981} Numerous model compounds and the large set of theoretical tools in one dimension made extensive comparisons between experiment and theory possible: for example, the universal scaling of spin excitations in the uniform spin-$\frac12$ Heisenberg chain was proposed theoretically and later confirmed experimentally.\cite{lake2005} A number of studies successfully extended the model by including interchain couplings and discussed the trends for the ordering temperature depending on the topology and magnitude of interchain couplings.\cite{schulz1996,yasuda2005,janson2009}

Alterations in the chain topology lead to a dramatic change in the magnetic properties. For example, there are several options to switch from the gapless spectrum of the uniform spin-$\frac12$ chain to a gapped spectrum. The latter offers an exciting opportunity to close the spin gap by an external magnetic field and to observe unusual phenomena, such as Luttinger liquid (LL) physics and the Bose-Einstein condensation of triplons in the gapless high-field phase.\cite{giamarchi2008} The simplest way to introduce a spin gap into a 1D system is to alternate the exchange couplings along the chain.\cite{johnston2000} Another option is the frustration of the chain by next-nearest-neighbor couplings.\cite{white1996} Finally, several chains can be joined into a spin ladder that shows a spin gap for an even number of legs.\cite{frischmuth1996} Despite the relatively simple chain geometries, such models are rather difficult to realize experimentally. There is still no experimental observation of the LL phase in the alternating spin-$\frac12$ chain, and experimental examples of gapped frustrated spin chains are rare.\cite{nilsen2008} The quest for spin-ladder systems was more successful. For example, recently a remarkable mapping of high-field properties onto the LL model in a (C$_5$H$_{12}$N)$_2$CuBr$_4$ compound was performed.\cite{klanjsek2008,ruegg2008,thielemann2009} 

Combining different features of the modified chain topology (alternation, frustration, and coupling into a ladder), one can achieve further interesting properties. For example, frustrated spin chains with alternating nearest-neighbor couplings are predicted to exhibit a magnetization plateau for a certain range of model parameters.\cite{totsuka1998} However, this prediction has never been tested experimentally due to the lack of proper model compounds. The problems with finding experimental realizations of certain spin models call for an alternative approach: the investigation of complex 1D models, stimulated by real materials. In the following we show that the recently discovered spin-$\frac12$ compound BiCu$_2$PO$_6$ closely corresponds to an interesting quasi-1D spin model combining all the three aforementioned features: frustration, spin-ladder geometry, and alternation of next-nearest-neighbor exchange couplings. 

Despite previous experimental and computational studies,\cite{kotes2007,mentre2009,kotes2010} the microscopic model of BiCu$_2$PO$_6$ remains controversial. To resolve this controversy, we apply a range of state-of-the-art computational techniques that reveal an accurate spin model and allow for a precise comparison with the experimental results. First, we analyze the crystal structure and outline the previous reports in Sec.~\ref{str}. After a brief description of the methods (Sec.~\ref{methods}), we proceed to extensive band structure calculations, derive a consistent spin model, and discuss the non-trivial implementation of this model in the crystal structure of BiCu$_2$PO$_6$ (Sec.~\ref{band}). In Sec.~\ref{experiment}, we report the magnetic susceptibility and the high-field magnetization measurements that challenge the proposed spin model and unambiguously measure the spin gap. Finally, we perform model simulations, investigate the microscopic physics of BiCu$_2$PO$_6$ at low energies (Sec.~\ref{diagonalization}) and in the presence of magnetic field (Sec.~\ref{sec:magn}), and conclude our study with a brief discussion and summary in Sec.~\ref{discussion}. 

\section{Crystal structure and magnetic properties}\label{str}
\begin{figure}
\includegraphics[scale=0.9]{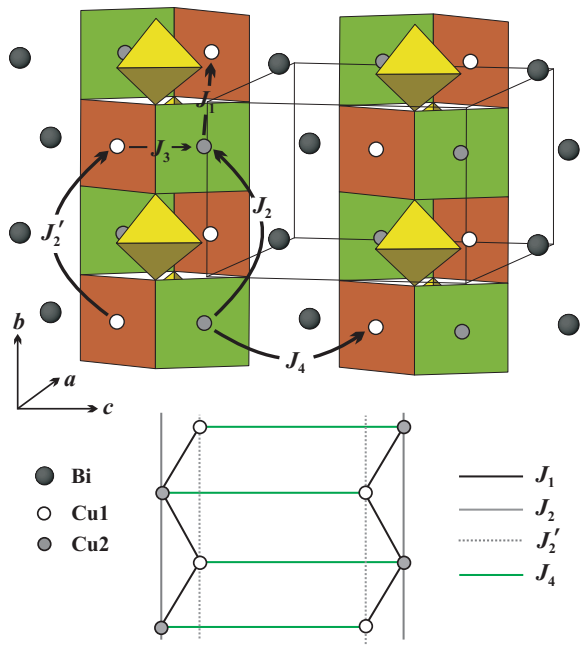}
\caption{\label{structure}
(Color online) Crystal structure of BiCu$_2$PO$_6$ with ribbons comprising CuO$_4$ plaquettes and PO$_4$ tetrahedra (top) and the spin model (bottom). Open and shaded circles denote the two inequivalent Cu positions, while the larger dark circles label the Bi atoms. More details on the structure are shown in Fig.~\ref{pathways}. The model is of the spin-ladder type and comprises four inequivalent couplings: the leg coupling $J_1$, the rung coupling $J_4$, and the frustrating next-nearest-neighbor leg couplings $J_2$ and $J_2'$. Note that the two legs of the ladder reside on \emph{different} structural ribbons.}
\end{figure}

The crystal structure of BiCu$_2$PO$_6$ (Fig.~\ref{structure}) shows pronounced 1D features with complex ribbons running along the $b$ direction.\cite{abraham1994} Each ribbon is formed by dimers of edge-sharing CuO$_4$ plaquettes. The plaquettes of the neighboring dimers share corners (oxygen sites), while the next-nearest-neighbor dimers are additionally connected by PO$_4$ tetrahedra. The spatial arrangement of the magnetic Cu atoms features both the spin-ladder and frustrated-spin-chain geometries (see Figs.~\ref{structure} and~\ref{pathways}). The stacking of the dimers reminds of the spin ladder with the leg coupling $J_1$ and the rung coupling $J_3$.\cite{note11} Yet, the interactions $J_1$ follow a zigzag pattern and form a frustrated spin chain, once the couplings between next-nearest neighbors are considered. The situation is further complicated by the two inequivalent Cu positions, leading to inequivalent next-nearest-neighbor couplings $J_2$ and $J_2'$.\cite{note12}

The complex crystal structure of BiCu$_2$PO$_6$ led to a controversy regarding the appropriate spin model of this compound. Koteswararao \textit{et al}.\cite{kotes2007} emphasized the spin-ladder feature of the structural ribbons and considered BiCu$_2$PO$_6$ as a system of $J_1-J_3$ ladders that are coupled by the inter-ribbon interaction $J_4$. This interpretation prevailed in further studies, focused on the effects of doping.\cite{bobroff2009,alexander2010,kotes2010,shiroka} However, band structure calculations, reported by the same authors,\cite{kotes2007} clearly showed sizable next-nearest-neighbor couplings $J_2$ and $J_2'$ that would inevitably frustrate the system.

Although similar at a first glance, Mentr\'e \textit{et al}.\cite{mentre2009} suggested a somewhat different spin model. Using inelastic neutron scattering (INS) and band structure calculations, they showed that the ladders are formed by the couplings $J_1$ and $J_4$, while the \emph{intra-ribbon} interaction $J_3$ is an \emph{inter-ladder} coupling. To fit the INS data, Mentr\'e \textit{et al}. also had to include the next-nearest-neighbor coupling $J_2$, but the difference between $J_2$ and $J_2'$ could not be resolved. 

Experimentally, BiCu$_2$PO$_6$ is a spin-gap material with a singlet ground state (no long-range ordering). The substitution of Cu by non-magnetic Zn atoms destroys the spin gap and leads to a spin freezing.\cite{bobroff2009,kotes2010} These features are fairly general and can be assigned to a range of simple 1D spin models (alternating chain, frustrated chain, two-leg ladder). However, the experimental data can not be described well by any of these models (see also Sec.~\ref{experiment}). The previous reports\cite{kotes2007,mentre2009} evidence the combination of the ladder-type geometry and the frustration by next-nearest-neighbor couplings. Yet, the precise way of this combination and, more importantly, the resulting physics remain unclear.

\section{Methods}\label{methods}
To evaluate the individual exchange couplings in BiCu$_2$PO$_6$, we performed scalar-relativistic density functional theory (DFT) band structure calculations using the full-potential local-orbital \texttt{FPLO} code (version 8.00-31).\cite{koepernik1999} The calculations were done in the framework of the local (spin) density approximation [L(S)DA], employing the exchange-correlation potential by Perdew and Wang.\cite{perdew1992} The symmetry-irreducible part of the first Brillouin zone was sampled by a mesh of 512 $k$-points for the crystallographic unit cell and 64 $k$ points for the supercells.

Superexchange couplings in insulating Cu$^{+2}$ compounds are intimately related to strong electronic correlations that cannot be properly treated within L(S)DA. To account for the correlation effects, we used two approaches. First, we mapped the half-filled LDA Cu $3d$ bands via an effective one-band tight-binding (TB) model onto a Hubbard model. Then, antiferromagnetic (AFM) exchange integrals were derived from the expression of the second-order perturbation theory. This procedure is referred below as the \emph{model approach}. In the second (supercell) approach, the correlation effects were treated in a mean-field approximation within the band structure calculations by applying the LSDA+$U$ method.\cite{note2} The on-site Coulomb repulsion parameter $U_d$ was varied in the $6-8$~eV range,\cite{johannes2006,schmitt2010,moeller2009,schmitt2009} while the on-site exchange parameter $J_d$ was fixed to 1~eV. Total energies for different types of collinear magnetic ordering were obtained within the crystallographic unit cell and the two supercells, doubled along the $b$ or $c$ directions. The calculated energies were mapped onto a Heisenberg model, and individual exchange couplings were derived. More details on the computational procedure are given in Sec.~\ref{band}.

The resulting spin model was compared to the experimental results from magnetic susceptibility and high-field magnetization measurements. Powder samples of BiCu$_2$PO$_6$ were prepared by firing a stoichiometric mixture of Bi$_2$O$_3$ (99.9~\% purity), CuO (99.99~\% purity), and NH$_4$H$_2$PO$_4$ (99.9~\% purity) in air. The mixtures were first annealed at 400~$^{\circ}$C for 10~hours and then at 850~$^{\circ}$C for 40~hours with one intermediate grinding. The resulting samples were single-phase, as confirmed by x-ray diffraction (STOE STADI-P diffractometer, CuK$_{\alpha1}$ radiation, transmission geometry). The magnetic susceptibility was measured in fields up to 5~T in the temperature range $2-700$~K using a Quantum Design MPMS SQUID magnetometer. 

High-field magnetization measurements were performed at Hochfeld-Magnetlabor Dresden at 1.4~K temperature in fields up to 60~T using a pulsed magnet. Details of the measurement technique are given in Ref.~\onlinecite{high-field}. 
%powered by a 1.44 MJ capacitor bank. With an inner bore of 20~mm the magnet yielded fields up to 60~T with a rise time of 7~ms and total pulse duration of about 20~ms. The magnetic moment of the sample was obtained by integrating the voltage induced in a compensated pick-up coil system surrounding the sample. The measurement was performed twice at 1.4~K temperature, with and without the sample, in order to remove the background signal contribution. 
The curves measured on increasing and decreasing field coincided, indicating the lack of any irreversible effects upon magnetization of the sample.

Thermodynamic properties of the BiCu$_2$PO$_6$ spin model were calculated by a full diagonalization for finite lattices with $N=16$ and $20$ sites and periodic boundary conditions. To obtain the low-energy excitations, we performed Exact Diagonalizations (ED) using the Lanczos algorithm that allowed to extend the system size up to $N=36$. The results are well converged with respect to the system size even for $N=16$ and $20$, thus the finite-size effects for the spin model under consideration are relatively small. To obtain the magnetization process of BiCu$_2$PO$_6$, we have used, in addition to ED, the Density Matrix Renormalization Group (DMRG)\cite{White1998,Schollwock2005} method with open boundary conditions with up to 128 rungs.
Further details are given in Secs.~\ref{experiment},~\ref{diagonalization}, and~\ref{sec:magn}.

\section{Derivation of the spin model}\label{band}
Spin models with exchange couplings derived from DFT have been previously reported in Refs.~\onlinecite{kotes2007} and~\onlinecite{mentre2009}. However, the analysis remains incomplete, since the two inequivalent next-nearest-neighbor couplings (between crystallographically different Cu sites) were considered to be equivalent. In the following, we apply two complementary approaches that evaluate all the relevant exchange integrals and establish the microscopic model. Additionally, we analyze in detail the structural features that cause the unusual implementation of the ladder-type spin lattice in BiCu$_2$PO$_6$. 

\subsection{LDA and model approach}
\begin{figure}[t]
\begin{center}
\includegraphics[width=0.9\linewidth,angle=0]{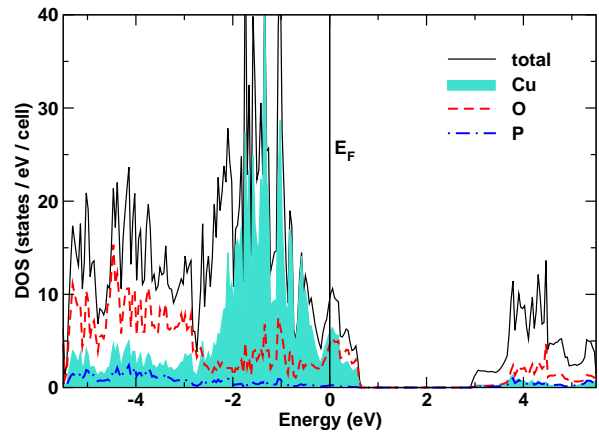}
\end{center}
\caption{\label{dos}(Color online) Total and site-projected DOS obtained from LDA. The vertical line at zero energy denotes the Fermi level $E_F$. The bands near the Fermi level primarily comprise Cu and O states. The shading in the plot denotes the Cu-$3d$ states, while
the dashed line represents the O $2p$ states.}
\end{figure}

\begin{figure}[t]
\begin{center}
\includegraphics[width=0.99\linewidth]{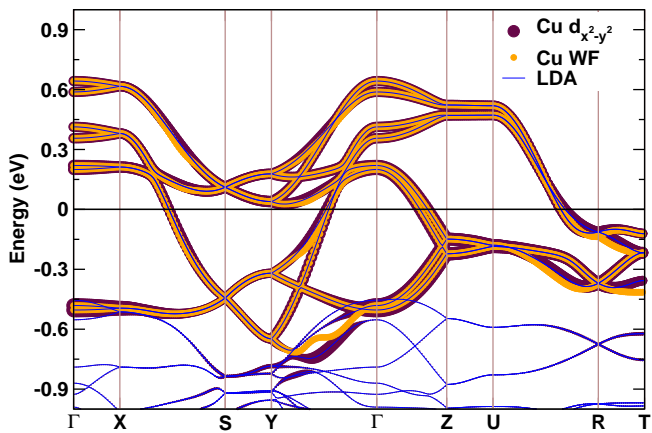}
\end{center}
\caption{\label{bands}(Color online) LDA band structure (thin blue lines), the WF-based fit of the tight-binding model (bright orange dots), and the contribution of the Cu $d_{x^2-y^2}$ orbital (dark purple dots). The high-symmetry $k$-path in terms of the reciprocal lattice parameters is as follows: $\Gamma(0,0,0)$, $X(0.5,0,0)$, $S(0.5,0.5,0)$, $Y(0,0.5,0)$, $\Gamma$, $Z(0,0,0.5)$, $U(0.5,0,0.5)$, $R(0.5,0.5,0.5)$, $T(0,0.5,0.5)$. The bands are highly dispersive along $X-S$, $Y-\Gamma-Z$, and $U-R$ which represent the leading interactions within the crystallographic $bc$ plane and the quasi-2D nature of the system.}
\end{figure}

Fig.~\ref{dos} shows the LDA density of states (DOS) of BiCu$_2$PO$_6$. The valence band spectrum is formed mainly by copper $3d$ and oxygen $2p$ orbitals, with a sizable contribution from phosphorous $3p$ orbitals below $-3$~eV. The states above $-0.6$~eV are formed by the Cu $3d_{x^2-y^2}$ orbital, in agreement with the expected ligand-field splitting.\cite{note1} The shapes and positions of the bands close to the Fermi level ($E_F$) are somewhat different from the $N^{\text{th}}$-order muffin-tin orbital (NMTO) result of Ref.~\onlinecite{kotes2007}, where the Cu $3d_{x^2-y^2}$ bands are separated from the lower-lying bands. This difference represents a known shortcoming of the NMTO method.\cite{rosner2009} To check our findings, we repeated the calculation using the full-potential code Wien2K. The resulting band structure is in excellent agreement to that from FPLO. Irrespective of the computational method, the LDA energy spectrum is metallic due to the underestimation of the correlation effects in this approximation. Experimentally, the green-colored BiCu$_2$PO$_6$ is a magnetic insulator. The insulating behavior is readily reproduced by the LSDA+$U$ calculations (see Sec.~\ref{lsda+u}).

\begin{table}[t]
\caption{\label{tbm} Leading hoppings of the tight-binding model and the resulting AFM exchange
couplings. The exchange pathways indicated in the first column are explicitly depicted in Figs.~\ref{structure} and~\ref{pathways}. The AFM part of the exchange integral is obtained by mapping the transfer integrals to an extended Hubbard model and eventually to a Heisenberg model using $J^{\AFM}_i=4t^2_i/U_{\eff}$ with $U_{\eff}=4.5$~eV.}  
\begin{ruledtabular}
\begin{tabular}{@{\extracolsep{\fill}}lllll}
Paths   & Cu--Cu distance & $t_i$   & Exchange  & $J^{\rm{AFM}}_i$  \\
%\cline{1-5}
 & (\AA) & (meV)  &  & (K)  \\
\hline
$t_{\rm{1}}$           & 3.21         & 146 & $J_{\rm{1}}$                 & 221  \\
$t_{\rm{2}}$           & 5.17 [Cu(2)] & 110 & $J_{\rm{2}} $                & 125  \\
$t_{\rm{2}}^{\prime}$  & 5.17 [Cu(1)] & 78  & $J^{\prime}_{\rm{2}}$        &  63  \\
$t_{\rm{3}}$           & 2.89         & 123 & $J_{\rm{3}}$                 & 157  \\
$t_{\rm{4}}$           & 4.91         & 140 & $J_{\rm{4}}$                 & 203  \\
%$t_{\rm{5}}$           &4.22 & 25 & $J_{\rm{5}}$                 & 7.3  \\
%$t_{\rm{5}}^{\prime}$  &4.43 & 20 & $J^{\prime}_{\rm{5}}$        & 4.6  \\
%$t_{\rm{6}}$           &5.81 & 27 & $J_{\rm{6}}$                 & 8.5  \\
%$t_{\rm{6}}^{\prime}$  &5.91 & 34 & $J^{\prime}_{\rm{6}}$        & 13   \\
\end{tabular}
\end{ruledtabular}
\end{table}

Eight Cu atoms in the crystallographic unit cell of BiCu$_2$PO$_6$ give rise to eight $3d_{x^2-y^2}$ bands (Fig.~\ref{bands}). We first fit these bands with a tight-binding model and extract the hopping parameters $t_i$ (Table~\ref{tbm}). The fitting procedure involves Wannier functions (WF) centered on Cu sites.\cite{eschrig2009} The application of the WF technique leads to a reliable fitting despite the slight overlap with the lower-lying bands. We are also able to resolve the couplings $J_2$ and $J_2'$ that correspond to the same Cu--Cu vector $(0,1,0)$ but refer to different Cu sites in the structure (Fig.~\ref{structure}). The hoppings are in agreement with the apparent features of the band structure. We find strong dispersion along the $\Gamma-Y$, $\Gamma-Z$, $X-S$, and $U-R$ directions which correspond to the crystallographic $bc$-plane with the couplings $J_1$, $J_2$, $J_2'$, $J_3$, and $J_4$. The dispersions along the other directions are less pronounced, indicative of a quasi-2D nature of this system. 

The hoppings are then introduced into a Hubbard model with the effective on-site Coulomb repulsion $U_{\eff}=4.5$~eV.\cite{johannes2006,nath2008,schmitt2010,moeller2009} In the limit of strong correlations ($t_i\ll U_{\eff}$) and in the half-filling regime, the low-lying excitations of the Hubbard model are described by a Heisenberg Hamiltonian comprising AFM exchanges $J_i^{\AFM}=4t_i^2/U_{\eff}$. The resulting $J_i^{\AFM}$ values are listed in Table~\ref{tbm}. The maximum long-range hoppings $t_l$ beyond $t_1-t_4$ amount to 30~meV, thus leading to $J_l^{\AFM}<10$~K. Since the leading exchange couplings amount to $150-250$~K, the minimal microscopic spin model can be restricted to five interactions: $J_1,J_2,J_2',J_3$, and $J_4$. 

A crucial fact to note at this juncture is the clear difference in the strengths of $J_2^{\rm{AFM}}$ and $J_2^{\prime\rm{AFM}}$. Geometrically, the hopping paths for these exchanges are rather similar (Fig.~\ref{pathways}), and this structural feature led the authors of Refs.~\onlinecite{kotes2007} and~\onlinecite{mentre2009} to assume $J_2=J_2'$.  In our analysis, we find that it is essential to treat these two exchanges independently, otherwise the band splittings at the $\Gamma$ point would not be reproduced correctly ({\it i.e.}, one obtains four doubly-degenerate bands with $J_2=J_2'$ instead of the eight separate bands). Hence the frustrating next-nearest-neighbor exchanges ``alternate'' along the $b$ axis (see Fig.~\ref{structure}) with $J_2'\simeq 0.5J_2$. A detailed analysis of this difference will be given in Sec.~\ref{cryst-chem}. 

\subsection{LSDA+$U$}
\label{lsda+u}
The model approach allows to estimate all the exchange couplings and to select the leading interactions for the minimum microscopic model. This is especially important for complex compounds with numerous and non-trivial superexchange pathways, like in BiCu$_2$PO$_6$. On the other hand, the model approach does not account for FM contributions that are relevant for short-range interactions.\cite{dioptase,schmitt2010} To correct the leading couplings for the FM contributions, we use the supercell approach. The \emph{total} exchange integrals, consisting of the FM and AFM contributions, are listed in Table~\ref{exchange} for the physically reasonable range of the $U_d$ values and for the two double-counting-correction (DCC) schemes. The latter is widely believed to be a minor feature of the LSDA+$U$ method, but our recent studies evidenced a sizable influence of the DCC on the exchange integrals in the case of short-range interactions.\cite{dioptase,cucl,cu2v2o7} 

The DCC is an essential part of the LSDA+$U$ approach, because a part of the on-site Coulomb repulsion energy is contained in LSDA and has to be subtracted from the total energy, after the explicit (mean-field) correction for the on-site Coulomb repulsion is included. The two most common corrections are around-mean-field (AMF)\cite{czyzyk1994} and fully-localized-limit (FLL).\cite{anisimov1993} For spin-$\frac12$ magnetic insulators, the difference between AMF and FLL was commonly believed to be minor.\cite{ylvisaker2009} By construction, FLL looks more appropriate for the strongly localized regime $t_i\ll U_{\eff}$.\cite{petukhov2003} Yet, both AMF and FLL readily reproduce the insulating state of BiCu$_2$PO$_6$. For example, we find the band gap $E_g\simeq 2.4$~eV and the magnetic moment of $0.81$~$\mu_B$ at $U_d=6$~eV in AMF.\cite{note6} FLL yields a somewhat lower gap $E_g\simeq 1.6$~eV at the same $U_d$ value, but the gap is readily increased up to 2.1~eV at $U_d=8$~eV. Experimental estimates of $E_g$ are presently lacking. However, even the experimental input will hardly resolve the ambiguity, since the $U_d$ value cannot be estimated precisely. Then, the exchange couplings should be analyzed in more detail.

\begin{figure*}
\includegraphics[scale=0.8]{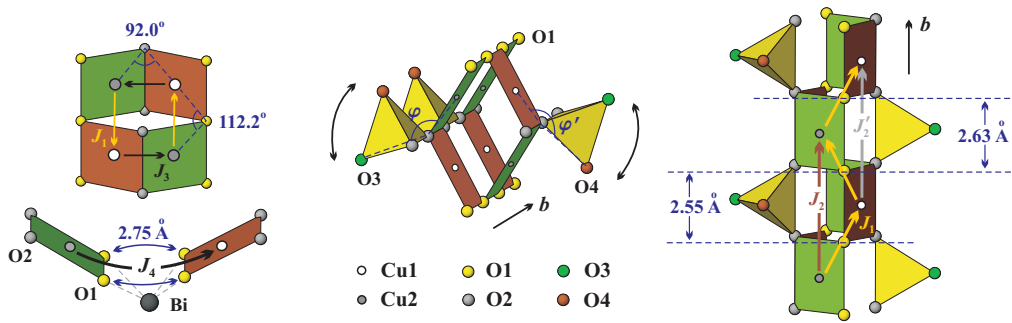}
\caption{\label{pathways}
(Color online) Parts of the crystal structure showing the details of individual superexchange pathways as well as the spin-ladder (top left panel) and the frustrated-spin-chain (right panel) features. The middle panel depicts the difference in the positions of the PO$_4$ tetrahedra for the couplings $J_2$ and $J_2'$. Curved arrows denote the rotations of the tetrahedra in the fictitious model structures (see text for details). The right panel shows the difference in the O1--O1 distances for $J_2$ and $J_2'$.
}
\end{figure*}

AMF and FLL produce similar estimates for most of the couplings: $J_1$, $J_2$, $J_2'$, and $J_4$ (see Table~\ref{exchange}). However, the short-range interaction $J_3$ is highly sensitive to the choice of the DCC. AMF suggests $J_3$ to be a weak coupling (either FM or AFM, depending on the $U_d$ value), while FLL ranks $J_3$ as one of the leading AFM couplings, comparable to $J_1$ and $J_2$. The FLL values essentially reproduce the previously published results by Mentr\'e \textit{et al.}\cite{mentre2009} that were also obtained within FLL but in a different band structure code. The model approach (Table~\ref{tbm}) evaluates $J_i^{\AFM}$, hence the FM contributions $J_i^{\FM}=J_i-J_i^{\AFM}$ can be calculated. Following this procedure, we find a simple microscopic argument that supports the AMF results with weak $J_3$. Both $J_1$ and $J_3$ arise from Cu--O--Cu superexchange with different angles at the oxygen atoms: $112.2^{\circ}$ and $92.0^{\circ}$, respectively (see the top left panel of Fig.~\ref{pathways}). According to the Goodenough-Kanamori rules,\cite{goodenough} the nearly $90^{\circ}$ superexchange of $J_3$ should yield the largest FM contribution. This conclusion conforms to the AMF results with $J_1^{\FM}=-45$~K and $J_3^{\FM}=-135$~K at $U_d=6$~eV. The FLL results are opposite, $J_1^{\FM}=-36$~K and $J_3^{\FM}=-16$~K. As $U_d$ is increased up to 8~eV, all the couplings are reduced, while the qualitative difference persists: $|J_3^{\FM}|>|J_1^{\FM}|$ in AMF, but $|J_3^{\FM}|<|J_1^{\FM}|$ in FLL.

\begin{table}[t]
\caption{\label{exchange} Total exchange couplings (in K) obtained from the LSDA+$U$ calculations. The $U_d$ value (in eV) denotes the Coulomb repulsion parameter of LSDA+$U$. The last column lists the double-counting correction scheme: around-the-mean-field (AMF) or fully-localized-limit (FLL).}
\begin{ruledtabular}
%\begin{tabular}{d@{\hspace{3em}}ddddd@{\hspace{3em}}c}
\begin{tabular}{r@{\hspace{1em}}rrrrr@{\hspace{1em}}c}
  $U_d$ & $J_1$ & $J_2$ & $J_2'$ & $J_3$ & $J_4$ &     \\\hline
  6     & 176   & 170   & 90     & 22    & 154   & AMF \\
  7     & 145   & 127   & 73     & -2    & 113   & AMF \\
  8     & 109   & 99    & 58     & -15   & 85    & AMF \\\hline
  6     & 185   & 166   & 93     & 141   & 243   & FLL \\
\end{tabular}
\end{ruledtabular}
\end{table}
The above considerations suggest the exchange couplings from AMF as a more reliable estimate for BiCu$_2$PO$_6$. For relevant examples from other compounds with a simpler magnetic behavior, we refer the reader to Sec.~\ref{cryst-chem}. Additionally, we note that computational results for $\beta$-Cu$_2$V$_2$O$_7$ (Ref.~\onlinecite{cu2v2o7}) and for several other Cu$^{+2}$-compounds\cite{cucl} also prefer AMF. Thus, we further rely on the AMF estimates and consider $J_3$ as a weak coupling. The low value of $J_3$ compared to $J_3^{\AFM}$ reduces the 2D $J_1-J_4$ model, obtained from the model approach, to a quasi-1D model, depicted in the bottom part of Fig.~\ref{structure}. This model basically follows the earlier proposal by Mentr\'e \textit{et al.}\cite{mentre2009} We find a two-leg spin ladder with the leg coupling $J_1$, the rung coupling $J_4$, and the next-nearest-neighbor frustrating couplings $J_2$ and $J_2'$ along the legs. Yet, there are two important differences to be emphasized. First, the two next-nearest-neighbor couplings are inequivalent and fairly different. The $J_2$ coupling connecting the Cu2 sites is twice as large as the coupling $J_2'$ between the Cu1 sites (see Tables~\ref{tbm} and~\ref{exchange}). Second, we can safely establish the quasi-1D nature of the spin model, because the $J_3/J_4$ ratio is below 0.2 (compare to $J_3/J_4=0.55-0.65$ in Ref.~\onlinecite{mentre2009}). Both results are very important for understanding the material. 

The difference between $J_2$ and $J_2'$ clearly alters the spin lattice. The pronounced one-dimensionality allows to simulate the behavior of the spin model on a quantitative level, despite the presence of the strong frustration that narrows the range of applicable simulation techniques. Before turning to the experiments and simulations (Sec.~\ref{experiment}), we will further discuss the non-trivial implementation of individual exchange couplings in the crystal structure of BiCu$_2$PO$_6$ and provide further support for the proposed spin model.

\subsection{Structural aspects of the magnetic exchange}
\label{cryst-chem}
The interactions $J_1$ and $J_3$ run between corner-sharing and edge-sharing CuO$_4$ plaquettes, respectively (top left panel of Fig.~\ref{pathways}). This geometry suggests Cu--O--Cu superexchange as the leading mechanism of the coupling and the angle at the oxygen atom as the key structural parameter determining the exchange integral. Following the Goodenough-Kanamori rules,\cite{goodenough} we find that $J_3$ with the Cu--O--Cu angle of $92.0^{\circ}$ is weakly AFM or even FM (see Table~\ref{exchange}). The pathway of $J_1$ reveals the sizably larger angle of $112.2^{\circ}$ and, consequently, a sizable AFM superexchange. A similar superexchange scenario is found in the mineral dioptase Cu$_6$Si$_6$O$_{18}\cdot 6$H$_2$O (green dioptase)\cite{dioptase} and, presumably, in its anhydrous counterpart (black dioptase). The spin lattice of dioptase comprises the AFM coupling $J_c$ between corner-sharing CuO$_4$ plaquettes (the Cu--O--Cu angle amounts to $107.6^{\circ}$ and $110.7^{\circ}$ for green and black dioptase, respectively) and the FM coupling $J_d$ between edge-sharing plaquettes ($97.4^{\circ}$ and $97.3^{\circ}$, respectively). More specifically, $J_c=78$~K and $J_d=-37$~K in green dioptase.\cite{dioptase} The nature of the exchange couplings in the dioptase lattice is confirmed by the magnetic structure that was directly investigated by neutron diffraction.\cite{wintenberger1993,belokoneva2002} Additionally, our recent computational study of green dioptase confirms the assignment of the exchange couplings and yields a consistent interpretation for all available experimental data.\cite{dioptase} The reference to the closely related superexchange scenario in dioptase should be taken as an additional argument for the weakness of $J_3$ and the resulting quasi-1D character of BiCu$_2$PO$_6$.

In fact, one can find further examples supporting the pronounced difference between $J_1$ and $J_3$. Numerous cuprates with chains of edge-sharing plaquettes are experimental realizations of frustrated spin chains with FM nearest-neighbor couplings. Such FM couplings arise from the Cu--O--Cu angle close to $90^{\circ}$ and typically range from $-100$~K to $-300$~K for oxide compounds (e.g., Li$_2$CuO$_2$, Li$_2$CuZrO$_4$).\cite{drechsler2007,lorenz2009} In BiCu$_2$PO$_6$, $J_3^{\FM}$ is smaller due to the folded arrangement of the plaquettes. Nevertheless, the pronounced FM contribution reduces the total exchange to a weak coupling, either FM or AFM, despite the sizable AFM contribution of $J_3^{\AFM}=176$~K (cf. Table \ref{tbm}). The leg coupling $J_1$ appears for the twisted configuration of corner-sharing plaquettes (see Fig.~\ref{pathways}) with the Cu--O--Cu angle of $112.2^{\circ}$. A similar configuration is found in AgCuVO$_4$, where the angle amounts to $112.7^{\circ}$, and a pronounced AFM exchange coupling $J\simeq 300$~K is found.\cite{moeller2009} Thus, our estimates of $J_1$ and $J_3$ are in line with the experience regarding other Cu compounds with firmly established microscopic models. 

All the above arguments support the quasi-1D model with weak $J_3$. In the following, we use this model as a working hypothesis to interpret the magnetic behavior of BiCu$_2$PO$_6$. The quasi-1D model captures the essential physics of the material, although certain features may require the extension of the model towards including $J_3$ or anisotropy effects (see Sec.~\ref{diagonalization} and~\ref{discussion}).

Taking $J_3$ as a weak interaction, we find $J_4$ to be the leading coupling along the $c$ direction. This coupling runs between the CuO$_4$ plaquettes of neighboring ribbons. The bonding between the ribbons arises from Bi cations (bottom left panel of Fig.~\ref{pathways}), yet Bi does not give any sizable contribution to the states near the Fermi level. Therefore, we assign $J_4$ to the Cu--O--O--Cu superexchange with the double O--O contact of 2.75~\r A. Similar couplings between the disconnected copper plaquettes have been reported for (CuCl)LaNb$_2$O$_7$ and Bi$_2$CuO$_4$.\cite{cucl,tsirlin2009,janson2007} Due to the large spatial separation of the Cu atoms (4.91~\r A), a sufficiently strong interaction arises for specific configurations of the ligand orbitals only (see Ref.~\onlinecite{tsirlin2009} for an instructive example). This explains the strong inter-ribbon coupling along the $c$ direction, in contrast to a very weak coupling between the structural ribbons along $a$ where the shortest Cu--Cu distance is 4.85~\r A.

Finally, we address the most puzzling feature of BiCu$_2$PO$_6$, the next-nearest-neighbor couplings $J_2$ and $J_2'$. While the other couplings can be tentatively assigned after a careful analysis of the superexchange pathways, the sharp difference between $J_2$ and $J_2'$ remains unexpected. The Cu--Cu distances for the two couplings are the same and amount to 5.17~\r A, the lattice parameter along the $b$ direction. On the other hand, $J_2$ and $J_2'$ correspond to different Cu positions and are \emph{inequivalent by symmetry}. Band structure calculations within the model and LSDA+$U$ approaches consistently suggest that $J_2'/J_2\simeq 0.5$ (see Tables~\ref{tbm} and~\ref{exchange}).

\begin{figure}
\includegraphics[width=0.99\linewidth]{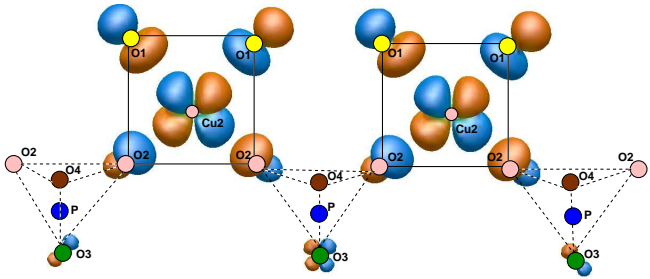}
\caption{\label{wannier}
(Color online) Wannier functions (``magnetic orbitals'') centered on Cu2 sites. Each orbital comprises the Cu $3d_{x^2-y^2}$ atomic orbital, large O1 and O2 $\sigma p$-contributions, and a smaller O3 $\sigma p$-contribution.}
\end{figure}

The couplings $J_2$ and $J_2'$ run between the copper plaquettes, joined by another plaquette via O1 and by a PO$_4$ tetrahedron via O2 (see the right panel of Fig.~\ref{pathways}). Thus, two different Cu--O--O--Cu channels are available. Despite the very similar Cu--O distances and Cu--O--O angles, there is a pronounced difference in the O1--O1 distances: 2.55~\r A for $J_2$ [the edge of the Cu1 plaquette] and 2.63~\r A for $J_2'$ [the edge of the Cu2 plaquette]. The shorter O1--O1 distance should lead to the stronger coupling $J_2$, in agreement with the computational result $J_2'<J_2$. At first glance, the O2 channel looks completely identical, because the O2--O2 distance is constrained by the edge of the PO$_4$ tetrahedron (2.56~\r A). Nevertheless, this channel also contributes to the difference between $J_2$ and $J_2'$. 

To get a deeper insight into the mechanism of the next-nearest-neighbor interactions, we inspect the Wannier functions for the Cu1 and Cu2 sites. Each WF comprises a Cu $3d_{x^2-y^2}$ orbital along with the $\sigma$-type $p$-orbitals of the neighboring oxygens O1 and O2 (Fig.~\ref{wannier}). We also find small, but significant, $\sigma$-contributions from second-neighbor oxygens O3 and O4 for the Cu2 and Cu1 WFs, respectively. These ``tail'' contributions arise from the specific orientation of the PO$_4$ tetrahedra: one of the O--O edges aligns along the Cu--O2 bond, i.e., the Cu2--O2--O3 ($\varphi$) and Cu1--O2--O4  ($\varphi'$) angles approach $180^{\circ}$. Indeed, we find $\varphi=140.4^{\circ}$ and $\varphi'=159.1^{\circ}$ in agreement with the smaller O3 contribution of about 1.0~\%, compared to 1.7~\% for O4.

\begin{figure}
\includegraphics[width=0.99\linewidth]{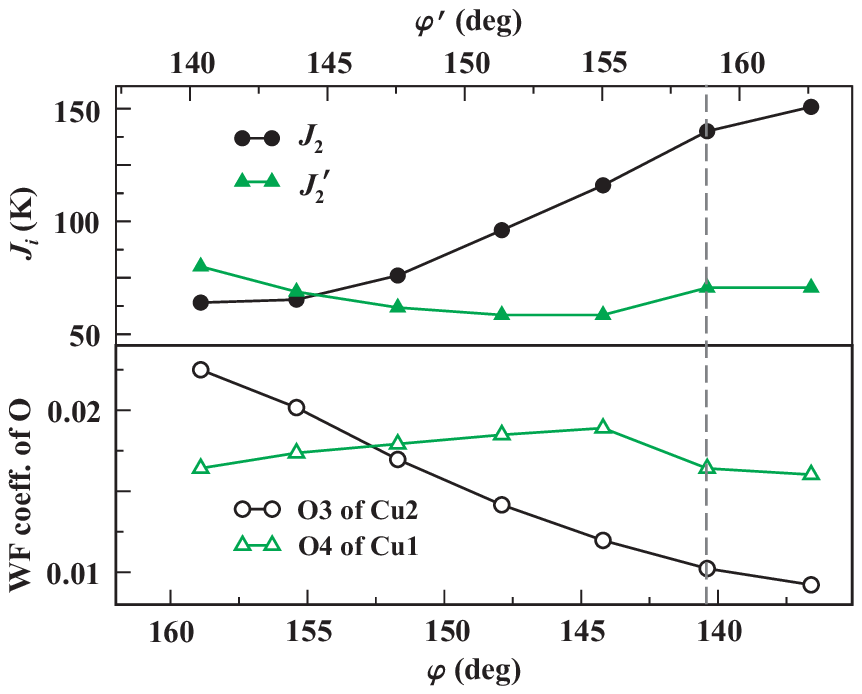}
\caption{\label{phi}
(Color online) Exchange integrals $J_2$ and $J_2'$ and the contribution of the second-neighbor oxygens (O3, O4) to the Wannier functions, depending on the position of the PO$_4$ tetrahedron. The dashed vertical line shows the angles in the BiCu$_2$PO$_6$ structure.}
\end{figure}

Although the tail features of the WFs look tiny, they have a strong effect on the exchange couplings. To probe this, we constructed fictitious model structures by rotating the PO$_4$ tetrahedra around the O2--O2 edge. Since the tetrahedra were kept rigid, only the $\varphi$ and $\varphi'$ angles were varied, while other geometrical parameters remained constant.\cite{note3} We found that the position of the tetrahedron leads to a dramatic change in the absolute values of $J_2$ and $J_2'$. As the $\varphi$ angle is increased towards $180^{\circ}$, the O3 contribution gets larger, and $J_2$ consequently decreases (Fig.~\ref{phi}). The rotation of the tetrahedra by $15^{\circ}$ makes $J_2$ and $J_2'$ equal, while the further rotation will switch the system to the $J_2'>J_2$ regime. The WF of Cu1 and the interaction $J_2'$ are less sensitive to the variation of the $\varphi'$ angle within the studied angle range.\cite{note4}

Our analysis shows that the structural features beyond the CuO$_4$ plaquettes have a sizable effect on the exchange couplings in Cu$^{+2}$ compounds. In BiCu$_2$PO$_6$, the tails of the WFs on the second-neighbor oxygens have $90^{\circ}$ orientation and should then reduce the AFM coupling (see Fig.~\ref{wannier}). This unexpected interference of the magnetic orbitals on the second-neighbor oxygen site is one of the microscopic reasons for the observed difference between $J_2$ and $J_2'$. It is worth noting that the role of non-magnetic side groups was emphasized theoretically long ago,\cite{khomskii1996} but is often not taken into account adequately in a quantitative description. Here, we have shown that the oxygen orbitals play the key role, while the phosphorous atom simply ``holds'' the four oxygens of the tetrahedron together. There is no appreciable phosphorous contribution at the Fermi level, and its contribution to the WF's is also minor (below 0.1~\%). This general mechanism, involving interacting oxygen atoms, has been recently found in vanadium phosphates\cite{tsirlin2010} and deserves further investigation in the compounds comprising other transition metals.

\section{Experimental results}
\label{experiment}
\subsection{Magnetic susceptibility}
\begin{figure}
\includegraphics{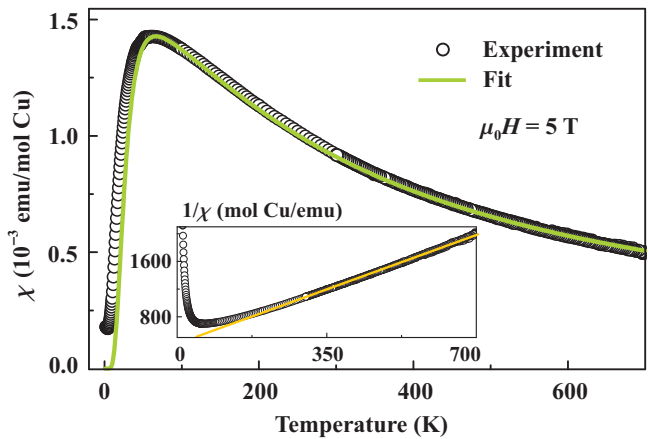}
\caption{\label{chi}
(Color online) Magnetic susceptibility of BiCu$_2$PO$_6$ measured in the applied field $\mu_0H=5$~T and the fit of the 1D spin model with $g\simeq 2.16$, $J_1\simeq 140$K, $J_2=J_1$, $J_2'=\frac12J_1$, and $J_4=\frac34J_1$ (simulation for a finite lattice with $N=20$~sites). The inset shows the Curie-Weiss fit above 300~K.}
\end{figure}

The temperature dependence of the magnetic susceptibility is shown in Fig.~\ref{chi} and resembles closely the data from Ref.~\onlinecite{kotes2007}. We find a broad maximum at $T_{\max}^{\chi}\simeq 62$~K, indicative of the predominantly AFM low-dimensional and/or frustrated behavior. The sharp decrease in the susceptibility below $T_{\max}^{\chi}$ is a signature of the spin gap. In the low-temperature region, the 0.1~T data show a weak upturn below 5~K. This upturn is largely suppressed in the field of 5~T and can therefore be assigned to a paramagnetic contribution of defects/impurities. Above 10~K, the susceptibility is field-independent in the studied field range $\mu_0H\leq 5$~T.

Above 200~K, the system approaches the Curie-Weiss regime. In order to improve previous studies,\cite{kotes2007,mentre2009} we measured the susceptibility at high temperatures up to 700~K and fitted the data above 300~K with the expression
\begin{equation}
\chi=\dfrac{C}{T+\theta}
\label{cw}
\end{equation}
where $\theta$ is the Curie-Weiss temperature and $C=N_A(g\mu_B)^2 S(S+1)/(3k_B)$ is the Curie constant. Our fit gives $C=0.447(1)$~emu K/mol~Cu, and $\theta=181(1)$~K (see the inset of Fig.~\ref{chi}). Fitting the data with an additional temperature-independent $\chi_0$ term leads to a small $\chi_0$, therefore, we neglect this term in further analysis. We establish the predominant AFM nature of the exchange interactions with an energy scale of about 200~K. The $C$ value corresponds to an effective moment of $1.89(1)$~$\mu_B$, slightly above the ideal spin-$\frac12$ value of $1.73$~$\mu_B$ and rather typical for Cu$^{+2}$ compounds.\cite{schmitt2010,janson2009} Since $T_{\max}^{\chi}/\theta\simeq 0.3$, strong frustration should be expected.

For further analysis, we fit the magnetic susceptibility using our microscopic spin model. Koteswararao \textit{et al.}\cite{kotes2007} have shown that the data do not conform to the model of isolated non-frustrated spin ladders. The introduction of interladder couplings does not significantly improve the description.\cite{note5} Therefore, realistic models with frustrating next-nearest-neighbor couplings have to be considered. Mentr\'e \textit{et al.}\cite{mentre2009} used a frustrated $J_1-J_2-J_4$ spin model and fitted the data with $J_1\simeq 140$~K, $J_2\simeq 0.5J_1$, and $J_4\simeq 0.4J_1$ (see also Ref.~\onlinecite{shiroka}), but this model did not take into account the difference between $J_2$ and $J_2'$. 

Here we employ the $J_1-J_2-J_2'-J_4$ model (Fig.~\ref{structure}) to fit the experimental magnetic susceptibility. This 1D frustrated spin model can be treated by exact diagonalizations for finite lattices or by renormalization-group techniques. The former turns out to be appropriate for the present problem due to the small finite-size effects and will be used here for the susceptibility fit. The unit cell comprises four inequivalent Cu$^{2+}$ ions, hence the number of sites $N$ in the finite cluster should be a multiple of four. To fit the experimental data, we first approximate our model by the following set of parameters $J_1=J_2$, $J_2'=\frac12J_2$, and $J_4=\frac34J_1$, according to Table~\ref{exchange}. The simulations yield the reduced susceptibility $\chi^*$ which can be fitted to the experimentally observed $\chi$ using
\begin{equation}
\chi=\dfrac{N_Ag^2\mu_B^2}{J_1}\chi^*
\end{equation}
with only two variable parameters: $g$ and $J_1$. The simulations for $N=16$ and $N=20$ sites provide almost identical susceptibility curves. Hence, finite-size effects are negligible and our simulations yield accurate results for the 1D spin model under consideration.

Our optimal fits yield $J_1\simeq 140$~K and $g\simeq 2.16$. The fitted $g$ value is typical for Cu$^{+2}$ compounds\cite{moeller2009,incuvo3} and also conforms to the effective magnetic moment of 1.89~$\mu_B$ which leads to $g=2.18$. The absolute value of $J_1$ is in remarkable agreement to the computational estimate of $100-150$~K (cf. Table~\ref{exchange}). The fit follows the experimental data down to 100~K (see Fig.~\ref{chi}). At lower temperatures, we find slight deviations from the experiment. For instance, the position of the susceptibility maximum $T_{\max}^{\chi}$ is overestimated and the theoretical curve lies slightly below the experimental data. This shows that our model overestimates the spin gap $\Delta$. We shall return to this issue below in Sec. \ref{diagonalization}. 

We also tried to vary the ratios of exchange integrals and found several fits of similar quality. In particular, the parameter set from Ref.~\onlinecite{mentre2009} ($J_1\simeq 140$~K, $J_2=J_2'\simeq 0.5J_1$, and $J_4\simeq 0.4J_1$) is also in agreement with the magnetic susceptibility data and yields a comparable $g=2.145$. However, this parameter set does not account for the difference between $J_2$ and $J_2'$. Since, as shown above, this difference is evidenced by two different computational approaches and has a clear structural origin, we regard the solution $J_1=J_2$, $J_2'=\frac12J_2$, and $J_4=\frac34J_1$ as the microscopically justified parameter set for BiCu$_2$PO$_6$.

\subsection{High-field magnetization and the spin gap}\label{magnetization}
The low-energy physics of BiCu$_2$PO$_6$ is characterized by the presence of a spin gap $\Delta$. Previous estimates of $\Delta$, based on the magnetic specific heat\cite{kotes2007,kotes2010} and Knight shift,\cite{bobroff2009,alexander2010} consistently suggested $\Delta\simeq 35$~K. The INS data revealed a smaller gap of 2~meV (about 23~K).\cite{mentre2009} The observed discrepancy calls for the application of further experimental methods, especially in light of the ambiguity of the specific heat and the Knight shift estimates, which arises from the fitting expressions that depend on the character of the spin excitations and, in particular, on the dimensionality of the system.

\begin{figure}
\includegraphics{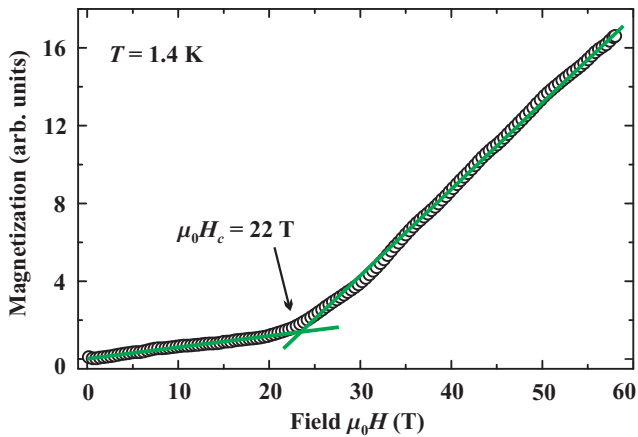}
\caption{\label{mvsh}
(Color online) Magnetization curve of BiCu$_2$PO$_6$ measured in pulsed field at $T=1.4$~K. The magnetization values are given in arbitrary units (a.u.) The arrow shows the critical field $H_c$ where the spin gap is closed. The solid lines are guides for the eye.}
\end{figure}

High-field magnetization data can provide a robust estimate of the spin gap. 
The magnetization process of BiCu$_2$PO$_6$ is presented in Fig.~\ref{mvsh}. At low fields, the magnetization shows a weak linear increase with the field until 
$\mu_0H_c\simeq 22$~T, where it bends upwards following a much steeper linear increase at higher fields.\cite{note8} The transition at $H_c$ implies the closing of the spin gap and can be used for the numerical estimate of $\Delta$. Similar to Ref.~\onlinecite{high-field}, we take $H_c$ as the point of the maximum curvature. We find $\Delta=g\mu_B \mu_0H_c/k_B \simeq 32$~K, in good agreement to the previous estimate $\Delta \simeq 35$~K obtained from the magnetic specific heat and the Knight shift data.\cite{kotes2007,kotes2010,bobroff2009,alexander2010}

The behavior of the magnetization for small fields needs to be discussed in more detail. In an ideal, SU(2) invariant and defect-free gapped system, the magnetization should be zero below $H_c$ (see also Fig.~\ref{mvsh_theory} below). Impurities give rise to a finite magnetization contribution, but this should typically saturate around 5~T at the present low temperature of 1.4~K. Since the measured magnetization keeps increasing up to $22$~T, we conclude that the weak linear field dependence for $H<H_c$ is due to the presence of weak anisotropic interactions in BiCu$_2$PO$_6$. One such anisotropy, which is known\cite{Miyahara} to give rise to a linear magnetization response in the gapped regime of similar ladder systems, is the Dzyaloshinksy-Moriya (DM) anisotropy.\cite{Dzya,Moriya} As explained in Ref.~\onlinecite{Miyahara}, an isolated AFM dimer with a DM energy term of the form $\vec{D}\cdot(\vec{S}_1\times\vec{S}_2)$ admixes triplet excitations into the singlet ground state, and this gives rise to a uniform magnetization response of the form $\vec{m}_u\propto\vec{D}\times(\vec{D}\times\vec{B})$ even far below the critical field. There is also a staggered response in first order in $D$ of the form $\vec{m}_s\propto\vec{D}\times\vec{B}$ which can be detected by a local probe, such as NMR experiments. Similar features arise in the spin-ladder Cu$_2$(C$_5$H$_{12}$N$_2$)$_2$Cl$_4$ compoud.\cite{Miyahara}
Hence, it is reasonable to expect that the linear response observed for BiCu$_2$PO$_6$ at $H<H_c$ stems from the presence of the DM anisotropy. 

For completeness, it is worth providing a brief discussion on the main DM vectors, based on the crystal symmetry of BiCu$_2$PO$_6$ (cf.~Fig.~\ref{structure}). First of all, a DM anisotropy on each rung is allowed by symmetry, since each rung comprises two inequivalent Cu sites and thus the inversion symmetry through the middle of each rung is lacking. The translational invariance along the $b$ axis (with a period of two rungs) necessitates that the DM vectors are the same on every second rung. Furthermore, the fact that the $ac$ plane is a reflection (i.e., crystallographic mirror) plane\cite{note7} confines the DM vectors to the $b$ direction. There is finally a screw axis symmetry along $b$ (translation along $b$ by one rung, followed by a C$_2$ rotation around the $b$ axis) which connects the sites of two consecutive rungs. This last symmetry necessitates that the DM vectors on the two consecutive rungs differ in sign. The DM terms are also expected for other, inter-rung couplings. 

Finally, we would like to point out that the measured magnetization data right above $H_c$ do not show any square root singularity (cusp) as is typical for 1D systems with a quadratic branch of magnetic excitations above the ground state (see also Fig.~\ref{mvsh_theory} below). This is probably related to the presence of the DM interactions mentioned above and the inter-ladder coupling $J_3$, which are both expected to smooth out the singularity. 

In Sec.~\ref{sec:magn} below, we provide a more detailed theoretical picture for the magnetization process, but first it is essential to understand the nature of the lowest magnetic excitations in BiCu$_2$PO$_6$. 

\section{Low-energy excitations from Exact Diagonalizations}\label{diagonalization}
%%%%%%%%%%%%%%%%%%%%%%%%%%%%%%%%%%%%%
\begin{figure}[!t]
\centering
\includegraphics[width=0.48\textwidth]{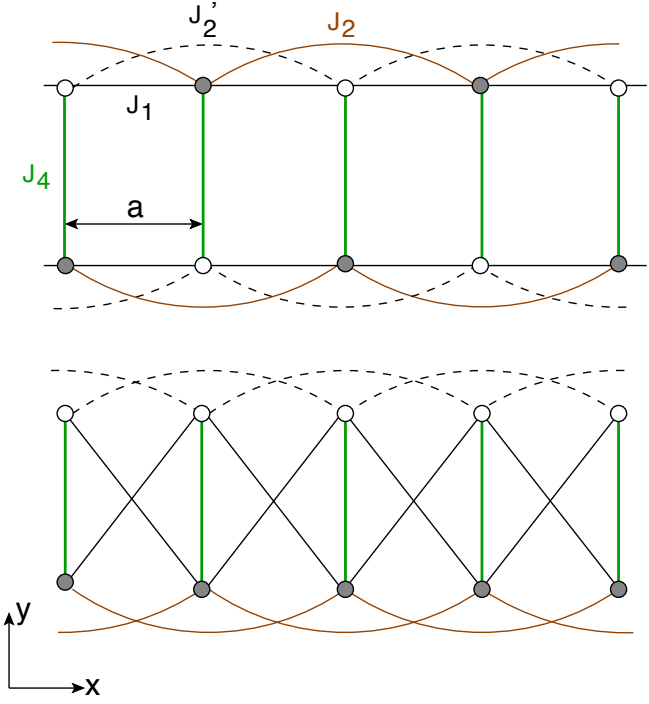}
\caption{(Color online) Top panel: The actual structure (disregarding the buckling) of the present model, and in the lower panel its topologically equivalent version 
obtained by flipping the two sites of every second rung.} 
\label{BothVersions}
\end{figure}
%%%%%%%%%%%%%%%%%%%%%%%%%%%%%%%%%%%%%
We have performed an exact diagonalization study of the model Hamiltonian discussed above (but without DM terms) with parameters $J_1=J_2=1$, $J_2'=0.5$, and $J_4=0.75$,   
using finite lattices of $N=12$, 16, 20, 24, 28, 32, and 36 sites with periodic boundary conditions along the legs ($x$-axis). The model is depicted in the upper panel of Fig. \ref{BothVersions}. Apart from translations along the legs by 2$a$, we also have two discrete spatial symmetries in this model. One is a reflection through any of the rungs ($\mc{P}_y$) and the other is a $\pi$-rotation ($\mc{C}_{2z}$) around the $z$-axis which is perpendicular to the plane of the ladder and passes through the center of a $J_1-J_4$ rectangle. Instead of the latter, we can take the generator consisting of a translation by $a$, combined with a reflection along the axis crossing the middle of all rungs. 

A first strong insight into physics of this model comes from a simple examination of the ground state expectation values of various local energy terms \mbox{$\langle\vec{s}_i\cdot\vec{s}_j\rangle$}. 
Owing to the spatial symmetries of the problem, there are four inequivalent bonds only. These are the bonds associated with the four different exchange couplings $J_1$, $J_2$, $J_2'$, and $J_4$ in each unit cell. The corresponding ground state expectation values, denoted as $e_1$, $e_2$, $e_2'$, $e_4$, are provided in Table \ref{bondstrengths} together with the total ground state energy per site $E/N=(2J_1e_1+J_2e_2+J_2'e_2'+J_4e_4)/2$. The latter shows only small finite-size variations for $N\ge 16$, which points to a very short correlation length.\cite{note9}
More importantly, we observe a sizably large value for the spin-spin correlations on the rungs, $e_4\simeq -0.47$, which is more than twice the values on the remaining bonds. 
This result tells us that the system is in the strong {\em rung} coupling regime, despite the fact that the leg couplings $J_1$ and $J_2$ are comparable to the rung coupling $J_4$.

\begin{table}[!b]
\caption{The ground state expectation values of the four different bond strengths $\langle\vec{s}_i\cdot\vec{s}_j\rangle$ per unit cell 
and the total energy per site $E/N$ in units of $J_1$.}
\begin{ruledtabular}
\begin{tabular}{c c c c c c}
$N$ & $e_1$ & $e_2$ & $e_2'$ & $e_4$ & $E/N$\\
\hline
12& -0.21568 & -0.17868 & -0.17162 & -0.42632 & -0.50780\\
16& -0.17466 & -0.19499 & -0.18411 & -0.47136 & -0.49494\\
20& -0.16633 & -0.21940 & -0.20928 & -0.46032 & -0.50096\\
24& -0.18409 & -0.18626 & -0.17460 & -0.47234 & -0.49800\\
28& -0.17210 & -0.20336 & -0.19237 & -0.47127 & -0.49860\\ 
32& -0.17683 & -0.19716 & -0.18571 & -0.47132 & -0.49858\\
%12& -0.21568016138634960 & -0.17867985200132200 & -0.17161948451657591 & -0.42632273626883332 \\
%16& -0.17466125275935829 & -0.19498637611213407& -0.18411269319936210 & -0.47136302418532250 \\
%20& -0.16632614210064223 & -0.21939725634087986 & -0.20927829043080076 & -0.46031567335210682 \\
%24& -0.18408747550550256 & -0.18626322983378538 & -0.17460294944633073 & -0.47233873304335283 \\
%28& -0.17210405412397009 & -0.20335770058467009 & -0.19237109786542339 & -0.47127339818759856 \\ 
%32& -0.17683481232229381 & -0.19715673981931578 & -0.18571223133539116 & -0.47131601576204013\\ 
\end{tabular}
\end{ruledtabular}
\label{bondstrengths}
\end{table}

\begin{figure}[!t]
\centering
\includegraphics[width=0.48\textwidth]{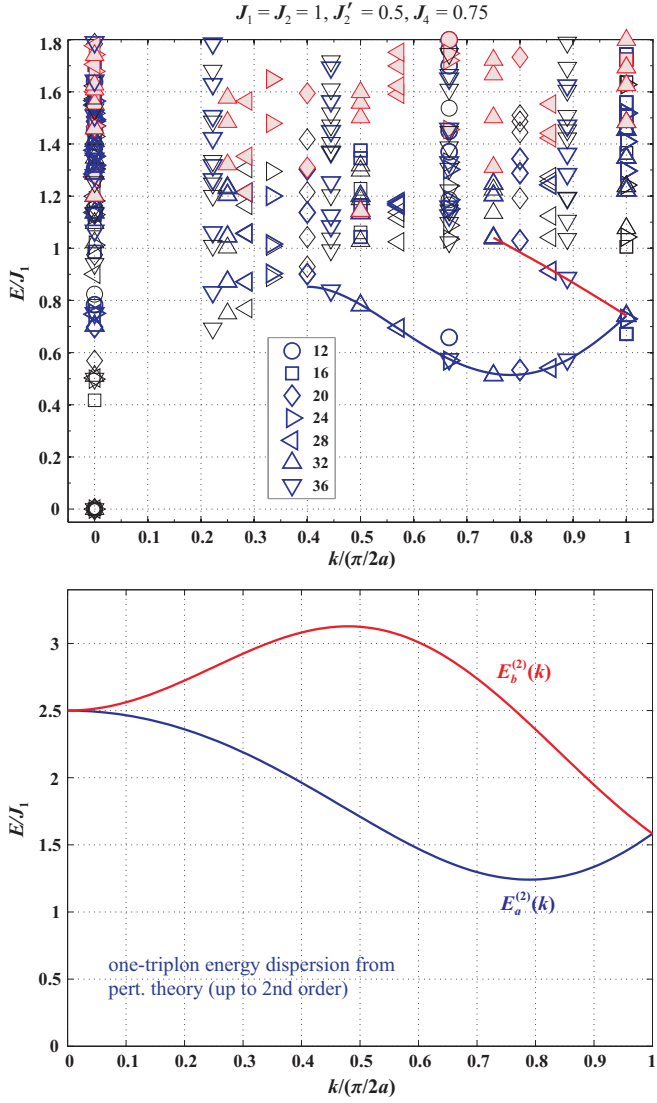}
\caption{(Color online) Top: Superimposed low-energy dispersions from exact diagonalizations on systems with $N=$12, 16, 20, 24, 28, 32, and 36 sites and for the parameters $J_1=J_2=1$, $J_2'=J_1/2$, and $J_4=0.75J_1$. 
Empty (black) symbols  denote the singlet $S=0$ states, thick open (blue) symbols denote the $S=1$ states, and filled (red) symbols denote the $S=2$ states.
The solid lines are polynomial fits to the visible parts of the lowest one-triplon excitation branches.
Bottom: The two one-triplon energy branches predicted from second-order perturbation theory around the strong coupling limit (cf. text). We emphasize here that the degeneracy of the two
bands at $k=0$ is an accidental feature of the second-order theory for the given values of the exchange parameters, while the degeneracy at $k=\pi/(2a)$ is a generic feature related to the fact that the model has a period $a$ and not 2$a$ along the legs of the ladder (cf. text).} 
\label{lespectraA}
\end{figure}

In Fig. \ref{lespectraA}, we have superimposed the low-energy excitations for each system size as a function of the allowed momentum quantum numbers so that we obtain a clear picture of the low-energy dispersion of the model. We observe that the lowest triplet (total spin $S=1$) excitations (thick open symbols) form a well-defined (coherent) branch separated from the 
continuum by a finite gap for $k\gtrsim 0.4\pi/(2a)$. This branch has an incommensurate minimum at $k_\textrm{min}\simeq 0.8\pi/(2a)$ at $\Delta^\text{ED} \simeq 0.5 J_1$. 
%One of the allowed momenta of the $N=20$ cluster sits very close to this minimum which explains why the ground state energy of this cluster is lower that the remaining $N\ge 16$ clusters. 
In addition to the lowest branch, we also find a second branch which is degenerate with the first at $k=\pi/(2a)$ but this shifts quickly to higher energies into the continuum for $k<\pi/(2a)$. 

%%%%%%%%%%%%%%%%%%%%%%%%%%%%%%%%%%%%%
%%%%%%%%%%%%%%%%%%%%%%%%%%%%%%%%%%%%%

Before we discuss the main implications of these results with regard to BiCu$_2$PO$_6$, we would like to provide a basic microscopic description of the excitation spectrum. 
To this end we perform a perturbative expansion around the limit of isolated rungs $J_1=J_2=J_2'=0$. 
We first introduce the singlet and triplet states of a single rung with sites 1 and 2 as $|s\rangle=\big( |\!\! \uparrow\downarrow\rangle-|\!\! \downarrow\uparrow\rangle \big) / \sqrt{2}$,  $|t_1\rangle=|\!\! \uparrow\uparrow\rangle$, $|t_{-1}\rangle=|\!\! \downarrow\downarrow\rangle$, and $|t_0\rangle=\big( |\!\! \uparrow\downarrow\rangle+|\!\! \downarrow\uparrow\rangle \big)/\sqrt{2}$. The unperturbed ground state is the product state of singlets on all rungs. 
Excitations arise by promoting one or more rungs into triplet states $|t_m\rangle$, with $m=\pm 1, 0$.  The inter-rung couplings have two effects. The first is that they renormalize the ground state energy as well as the energies in the one-triplon sector. The second is that they induce a finite amplitude for nearest-neighbor and next-nearest-neighbor hoppings of triplons in the one-triplon sector. Including the amplitude from all different processes and exploiting the translational invariance by 2$a$, one finds two separate bands of one-triplon excitations 
due to the fact that we have two rungs per unit cell in the model. Their energies relative to the renormalized ground state energy are given by 
$E_{\alpha,\beta}^{(2)}(k)=A_k\pm \left| B_k \right|$, with
\bea
A_k&=& J_4 + \frac{12 J_1^2+3(J_2+J_2')^2-4(J_2-J_2')^2}{16 J_4} \nonumber\\
&+&\left( \frac{J_2+J_2'}{2}+\frac{(J_2-J_2')^2-2J_1^2}{8 J_4} \right) \cos k \nonumber\\
&-&\frac{(J_2+J_2')^2}{16 J_4} \cos 2k,\nonumber\\
B_k&=& \frac{J_1}{2}(1+e^{-i k}) -\frac{J_1 (J_2+J_2') }{8 J_4} \nonumber\\
&&\times\left( 1+e^{-i k}+e^{i k}+e^{-2ik} \right)\nonumber
\eea
These second-order dispersions are shown in the lower panel of Fig.~\ref{lespectraA}. 
Although its prediction for the spin gap is more than twice higher than the exact value (shown in the upper panel), the second-order perturbation theory captures well the position of the minimum and the overall shape of the dispersion. 

%%%%%%%%%%%%%%%%%%%%%%%%%%%%%%%%%%%%%
\begin{figure}[!t]
\centering
\includegraphics[width=0.48\textwidth]{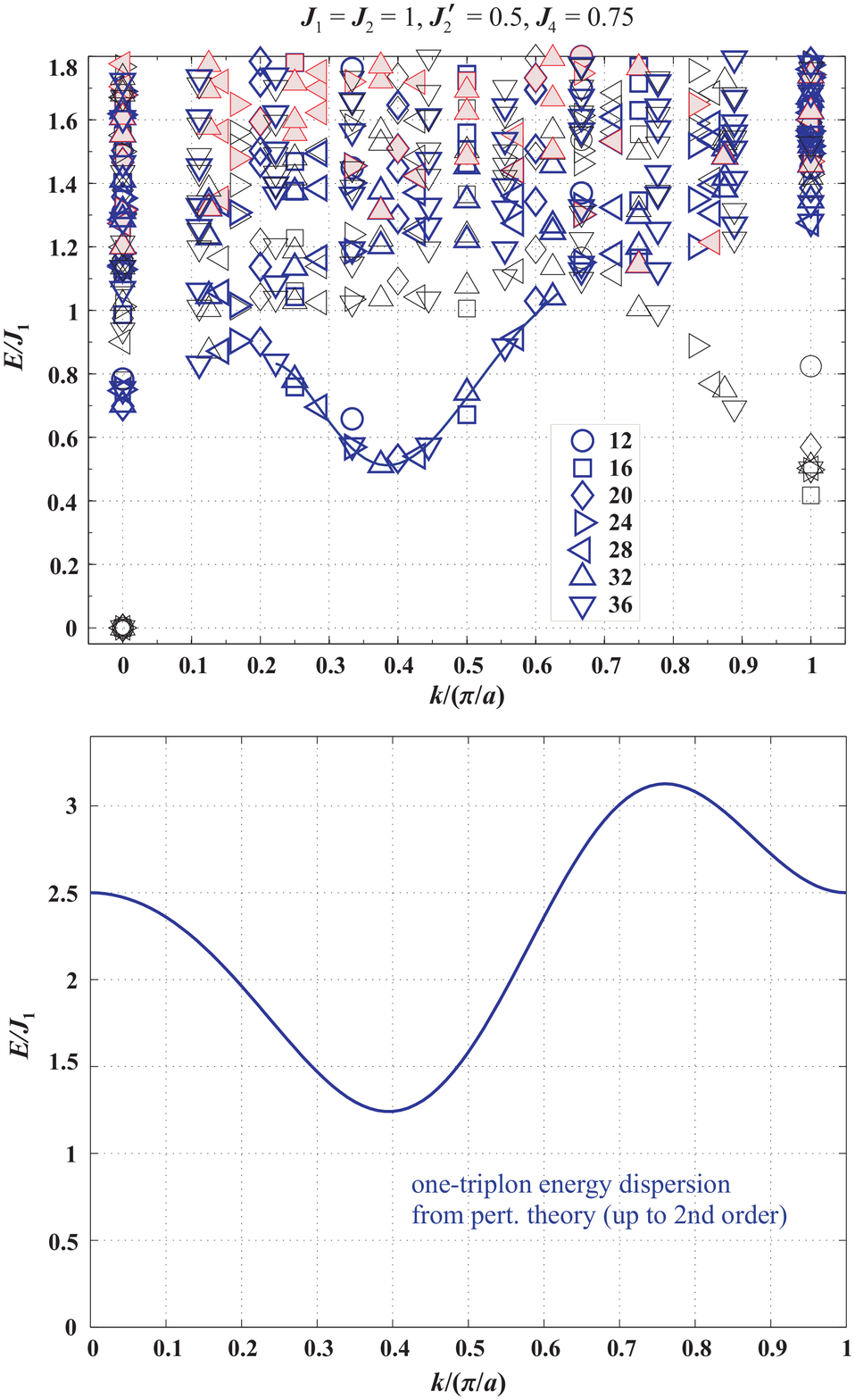}
\caption{(Color online) 
Top: Same as in Fig. \ref{lespectraA} but in the symmetry setup of the lower panel of Fig. \ref{BothVersions}. 
The solid line is a polynomial fit to the visible part of the lowest one-triplon excitation branch.
Bottom: The one-triplon energy dispersion predicted from second order perturbation theory around the strong coupling limit (cf. text).} 
\label{lespectraB}
\end{figure}
%%%%%%%%%%%%%%%%%%%%%%%%%%%%%%%%%%%%%

Next, we would like to comment that the degeneracy of the two branches at $k=0$ is not a generic feature of the exact dispersions but an accidental feature of the second-order expression for the given values of the exchange integrals. In higher orders of perturbation theory or for slightly different parameter values, this degeneracy will be lifted. 
In contrast, the degeneracy of the two branches at $k=\pi/(2a)$ is a generic feature and persists to all orders as seen in the exact spectra. The reason behind this is the presence of the discrete symmetry generator mentioned above (translation by $a$ followed by a reflection through the middle of all rungs). To see this, we may start from the upper panel of Fig.~\ref{BothVersions} and exchange the two sites of every second rung without altering the topology of the model. This gives the equivalent model shown in the lower panel of Fig.~\ref{BothVersions} which has period $a$ and not 2$a$. 
To elucidate this point, we may repeat the strong-coupling expansion in this alternative symmetry framework. To this end, we must take into account the  
extra negative signs that arise from the antisymmetry of the singlet rung wavefunction when flipping the two sites of every second rung. 
In terms of the new momenta, we now obtain a single one-triplon excitation band with energy dispersion
\bea
E(k)&=&J_4 + \frac{12 J_1^2+3(J_2+J_2')^2-4(J_2-J_2')^2}{16 J_4} \nonumber\\
&+& c_1 \cos k+c_2 \cos 2k + c_3 \cos 3k + c_4 \cos 4k
\eea
where
$c_1=-J_1+\frac{J_1 (J_2+J_2')}{4J_4}$, $c_2=\frac{J_2+J_2'}{2}+\frac{(J_2-J_2')^2}{8J_4}-\frac{J_1^2}{4J_4}$, $c_3=\frac{J_1 (J_2+J_2')}{4J_4}$, and $c_4=-\frac{(J_2+J_2')^2}{16J_4}$.
This dispersion is shown in the lower panel of Fig.~\ref{lespectraB}. It is clear that by folding this back into the Brillouin zone $[ -\pi/2a, \pi/2a ]$ we shall obtain the two branches shown before in the lower panel of Fig.~\ref{lespectraA}. It is also evident in this representation that the incommensurate nature of the dispersion arises already in first order and is dictated by the frustrated couplings $J_2$ and $J_2'$ which appear in the leading term in the above expression for $c_2$. 

For completeness, we present the exact diagonalization results in the new symmetry setup in the upper panel of Fig.~\ref{lespectraB}. 
Again, the overall shape of the lowest dispersion and the position of the minimum are in agreement with the prediction of the strong coupling expansion shown in the lower panel. 

An interesting feature which becomes better visible in the representation of Fig.~\ref{lespectraB} is the presence of a number of low-lying singlets for momenta close to $k=\pi/a$. 
These excitations can be understood as a singlet bound state of two triplons. Such singlet excitations could be captured by optical experiments, such as phonon-assisted 
infrared absorption. Singlet bound states of two triplons have been observed using such techniques in cuprate ladders.\cite{windt2001} In a broader context, the low-lying singlet at $k=\pi/a$ can be considered as a singlet mode going soft at the transition to a dimerized phase with dimers forming along the legs.\cite{vekua2007} In the model considered, this scenario might occur as the rung coupling $J_4$ is reduced further.

Let us now discuss the implications of the above findings for BiCu$_2$PO$_6$. 
Taking $J_1\simeq 140$~K from the fit of the susceptibility we obtain for the spin gap $\Delta^\text{ED}\simeq 0.5 J_1 \simeq 70$~K which is almost twice the value obtained from 
the high-field magnetization data, or the value reported by other groups.\cite{kotes2007,kotes2010,bobroff2009,alexander2010}
Hence, we find that the present model of an isolated frustrated ladder overestimates the value of the spin gap in BiCu$_2$PO$_6$, 
a fact that was already suggested from the behavior of the susceptibility at low temperatures.  One way to account for this discrepancy is to include a finite interladder coupling $J_3$. 
Along the lines of the previous perturbative analysis, one finds that $J_3$ gives rise to a first-order hopping of triplons along the $y$ direction. 
As a result, the two bands attain a common extra dispersion term of the form $-(J_3/2)\cos k_y$ (with $k_y$ in units of $\pi$ divided by the inter-ladder distance).
This shifts the minimum of the lowest band down by $J_3/2$. Thus, to account for the 35~K spin gap one would need an interladder coupling of the order of $J_3\simeq 70$K in this
simple approximation. However, in the present regime we expect the perturbative calculation to be only qualitatively correct, so that a precise determination of the interladder coupling either needs to come from more elaborate theoretical approaches (such as density matrix renormalization group simulations of coupled ladders) or, ultimately, from inelastic neutron scattering experiments on single crystals.

\section{Magnetization process from ED and DMRG} \label{sec:magn}
\begin{figure}
\includegraphics[width=0.48\textwidth]{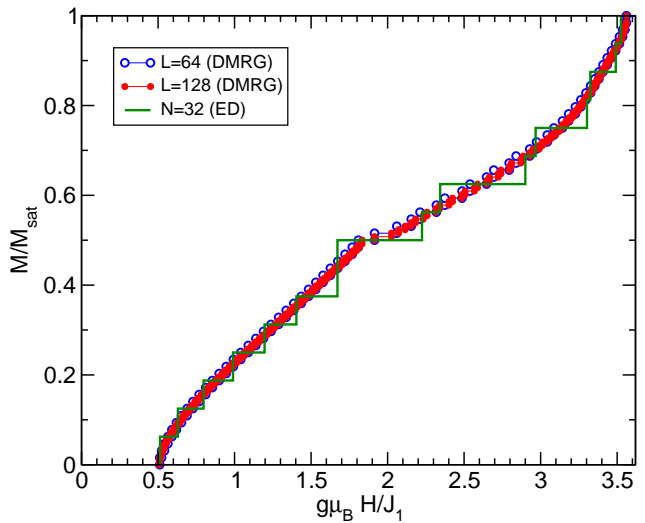}
\caption{\label{mvsh_theory}
(Color online) Magnetization curve of BiCu$_2$PO$_6$ as obtained from DMRG and ED.}
\end{figure}

Here we revisit the magnetization process of BiCu$_2$PO$_6$, in the light of the physical picture obtained above for the lowest magnetic excitations. 
To this end, we have employed Lanczos diagonalizations up to $N=32$ sites with periodic boundary conditions, as well as DMRG simulations with up to $L=128$ rungs using open boundary conditions.  
Some representative magnetization curves are shown in Fig.~\ref{mvsh_theory}. 
The results from the two largest clusters treated by DMRG ($L=64, 128$ rungs) converge to a rather smooth magnetization curve. 
They also give a critical field $H_c$ almost identical to the one obtained from ED for 32 sites, 
which further corroborates the value of the spin gap $\Delta^{\text{ED}} \simeq 0.5 J_1$ given above. 

To discuss the nature of the magnetization process in more detail, we distinguish three different regimes, namely the one at low magnetizations above $H_c$, the one at high magnetizations as we approach the saturation field $H_\text{sat}$, and the intermediate regime. The low magnetization regime can be qualitatively understood on the basis of gradually filling the excitation band of Fig.~\ref{lespectraB} (bottom) with triplons as we ramp up the field above $H_c$. One immediate consequence is the presence of a square-root singularity in the magnetization right above $H_c$ (cf. Fig.~\ref{mvsh_theory}) which is due to the quadratic dispersion above the minimum. Another important ingredient in this consideration is the presence of two incommensurate minima (at $k\simeq \pm 0.4 \pi/a$) in the triplon dispersion which, given the local hard-core constraint of the triplons, gives four Fermi points. Thus if the four-Fermi-point fix point is indeed stable, the effective low-energy theory of BiCu$_2$PO$_6$ at low magnetizations is a two-component LL. 

\begin{figure}
\includegraphics[width=0.48\textwidth]{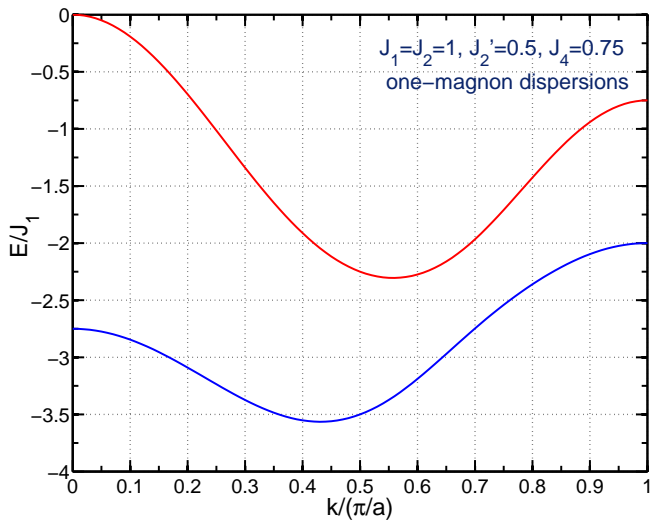}
\caption{\label{onemagnon}
(Color online) One-magnon energy dispersions obtained analytically (see Eq.~(\ref{Eq:onemagnon})).}
\end{figure}

In a similar way, the magnetization process close to the saturation field can be understood starting from the fully polarized state and gradually filling the one-magnon excitation bands by single spin flips. Using the setup of the lower panel of Fig.~\ref{BothVersions} and setting the energy of the fully polarized state to zero, one obtains two one-magnon bands which are given by the eigenvalues of the matrix 
\be\label{Eq:onemagnon}
\mc{H}_{\text{one-magn}}=
\left(\begin{array}{cc}
u_k & v_k \\
v_k & u_k'
\end{array}\right)
\ee
with $u_k=-(J_4/2+J_1+J_2)+J_2 \cos 2k$, $u_k'=-(J_4/2+J_1+J_2')+J_2' \cos 2k$, and $v_k=J_4/2+J_1 \cos k$.
The two one-magnon bands are shown in Fig.~\ref{onemagnon}. As expected, we find that each band has two minima at incommensurate wavevectors. In particular, the minima of the lowest band sit at $k=\pm 0.43131\pi/a$, which are close to the minimum $k$-points of the triplon dispersion of Fig.~\ref{lespectraB} (bottom). The corresponding minimum energy $E_{\text{min}}=-3.5643 J_1$ gives the saturation field $H_{\text{sat}}=3.5643 J_1/(g\mu_0\mu_B)$, in agreement with the numerical results of Fig.~\ref{mvsh_theory}. By gradually filling the minimum of the lowest one-magnon branch, we describe the magnetization process as we decrease the field below $H_{\text{sat}}$.
Similar to the low magnetization regime, the quadratic dispersion around the one-magnon minimum gives rise to a square-root singularity right below $H_{\text{sat}}$ which can be seen in our numerical results of Fig.~\ref{mvsh_theory}. In addition, the presence of two ``incommensurate'' minima (at $\pm 0.43131 \pi/a$) in the lowest one-magnon branch opens a possibility that 
the appropriate low-energy effective theory of BiCu$_2$PO$_6$ at high fields is a two-component LL.

It is presently unclear whether the possible two-component LL phases discussed at low and high fields form a single phase, or whether they are separated by one or 
more intervening phases at intermediate magnetizations. Inspecting the numerical results displayed in Fig.~\ref{mvsh_theory}, a plateau might, for instance, occur 
at \mbox{$M=M_\mathrm{sat}/2$}. The phase immediately above the plateau also requires further investigation 
since there is a possibility for a one-component LL phase before we reach the high-field two-component LL phase.
Such a rich interplay between one- and two-component LL phases and plateaux is realized in the frustrated antiferromagnetic $J_1-J_2$ Heisenberg chain model (see, e.g., Ref.~\onlinecite{Hikihara} and references therein). Testing and confirming the scenario outlined here for the physics of BiCu$_2$PO$_6$ in high magnetic fields requires a separate and more detailed investigation which is, however, beyond the scope of this article.

Let us finally compare to the experimental magnetization data of Fig.~\ref{mvsh}. Given our earlier estimate of $J_1\simeq 140$~K from the susceptibility fit, we obtain $H_\text{sat}\simeq 345$~T, 
which is much larger than the range of fields accessible in our experiment ($H_\text{max}=60$~T). Hence, the highest magnetization values reported in Fig.~\ref{mvsh} correspond to less than 10\% of $M_\text{sat}$. In contrast to the above theoretical predictions, the measured magnetization does not show any square-root singularity right above $H_c$. 
As we discussed in Sec.~\ref{magnetization}, this gives evidence for inter-ladder coupling $J_3$ and/or DM interactions which smooth out the singularity. 

\section{Discussion and conclusions}
\label{discussion}
Using DFT band structure calculations, we derived the minimum microscopic model of BiCu$_2$PO$_6$. This model is based on a two-leg-ladder lattice and comprises four antiferromagnetic exchange couplings: $J_1$ along the legs, $J_4$ along the rungs, and the frustrating next-nearest-neighbor couplings $J_2$ and $J_2'$ along the legs (Fig.~\ref{structure}). Although such a model does not provide a complete and quantitative description of the compound, it is a reasonable compromise between the complexity of the system and the capabilities of present-day numerical simulation techniques for the evaluation of ground-state and finite-temperature properties of frustrated quantum spin systems. We showed that the ladder geometry leads to strong spin correlations on the rungs, despite the sizable frustration and the weaker rung coupling. This feature might explain why the simple model of the unfrustrated spin ladder reproduces certain properties of BiCu$_2$PO$_6$, especially the behavior upon the chemical substitution.\cite{bobroff2009,alexander2010,shiroka} On the other hand, the reduction to the simple ladder model cannot be justified microscopically, since the frustrating coupling $J_2$ is of the same order as the leg and the rung couplings $J_1$ and $J_4$, respectively. In particular, the coupling $J_2$ has an effect on the spin gap. The simple $J_1-J_4$ two-leg ladder with $J_4=\frac34J_1$ shows a spin gap of about $0.3J_1$,\cite{johnston1996} while in our model the gap amounts to $0.5J_1$. Thus, the frustration enhances the gap in a spin ladder, similar to a conventional frustrated spin chain.\cite{white1996}

We interpret BiCu$_2$PO$_6$ as a system of two-leg ladders with frustrating couplings along the legs. The absolute values of individual exchange couplings leave an ambiguity to describe the system as a frustrated spin ladder or as coupled frustrated spin chains. Indeed, the actual system shows features of both models. On the one hand, the strongest correlations are found on the rungs, as in ordinary ladders. On the other hand, the correlations along the legs are incommensurate and lead to the spin gap, being minimal at an incommensurate position in the Brillouin zone.

BiCu$_2$PO$_6$ is a peculiar spin-ladder system interesting for future investigation. One of the exciting branches could be high-field studies above $H_c$. Recent experiments on (C$_5$H$_{12}$N)$_2$CuBr$_4$ evidenced the emergence of the LL physics in the high-field phase of the two-leg spin ladder.\cite{klanjsek2008,ruegg2008} BiCu$_2$PO$_6$ offers an opportunity to explore similar effects in the presence of the frustration, where the incommensurate position of the gap might lead to a two-component LL or instabilities thereof at fields just above $H_{c}$. Another advantage is the relative ease of the chemical substitution that has stimulated a range of experimental studies on Zn- and Ni-substituted samples.\cite{bobroff2009,alexander2010,shiroka} Here, again, the incommensurate leg spin-spin correlations could influence the effective interaction mediated between the impurity-induced localized spins, and thus lead to hitherto unobserved frustration effects.

While working on the minimal microscopic model, one also has to understand its limitations. The main and most severe limitation is the reduction to a purely 1D regime by neglecting $J_3$. In fact, our band structure calculations suggest $J_3/J_4<0.2$, i.e., $|J_3|\leq 25$~K. If we adjust $J_3$ to account for the actual spin gap $\Delta\simeq 32$~K $\simeq 0.2J_1$, a larger value is obtained (see Sec.~\ref{diagonalization}). Additionally, the shape of the magnetization curve with the linear increase right above $H_c$ (Sec.~\ref{magnetization}) may exclude a purely 1D scenario and point to sizable inter-ladder couplings. Considering all these arguments, we conclude that the inter-ladder coupling $J_3$ is likely relevant for the full picture, but its accurate estimate remains a challenging task. Band structure calculations equally allow for FM or AFM $J_3$ (Table~\ref{exchange}). Experimental estimates would require theoretical information on a complex 2D $J_1-J_2-J_2'-J_3-J_4$ frustrated spin system with long-range couplings $J_2$ and $J_2'$. Such a system is basically beyond the capabilities of present-day numerical methods. Therefore, the most reasonable approach could be analytical perturbation treatment, based on the accurate results for the 1D model. We believe that this approach will help to clarify the complex magnetic behavior of BiCu$_2$PO$_6$ and to improve the theoretical estimate of the spin gap with respect to the experimental value $\Delta\simeq 32$~K.

The second limitation of our model is the lack of anisotropy effects. In particular, the DM interactions scale with $J$ and can be sizable due to the strong isotropic exchange of $100-150$~K. The DM couplings are allowed for all the bonds of the spin lattice with few restrictions on the arrangement of the $\Dv$ vectors with respect to the crystal axes (see also Sec.~\ref{magnetization}). The comprehensive investigation of the anisotropy effects would require electron spin resonance measurements on single crystals along with sophisticated band structure calculations. Presently, we note that the increase in the magnetization below $H_c$ (Fig.~\ref{mvsh}) is a possible signature of the DM couplings. The non-zero Knight shift at low temperatures\cite{bobroff2009,shiroka} may have the same origin.

In summary, our study provides a comprehensive description of isotropic exchange couplings in the spin-$\frac12$ quantum magnet BiCu$_2$PO$_6$. We interpret this compound as a two-leg spin ladder with frustrating next-nearest-neighbor couplings along the legs. The leg coupling ($J_1$), the rung coupling ($J_4$), and one of the next-nearest-neighbor couplings ($J_2$) amount to $120-150$~K, while the other next-nearest-neighbor coupling $J_2'$ is half of $J_2$ due to the subtle structural differences between the respective superexchange pathways. The complex crystal structure of the compound leads to a non-trivial implementation of the spin ladder with two legs residing on different structural ribbons. The proposed spin model is a derivative of the simple two-leg spin ladder and shows leading spin correlations on the rungs. Frustrating couplings increase the spin gap and induce the incommensurate minimum of the triplon dispersion as well as an exotic behavior in high magnetic fields. The effects beyond our spin model include the inter-ladder coupling and the anisotropy. Experimental data show possible signatures of these effects and call for further investigation of BiCu$_2$PO$_6$ by means of inelastic neutron scattering and electron spin resonance measurements on single crystals.

\acknowledgments
We are grateful to Walter Schnelle for high-temperature susceptibility measurements and for careful reading of the manuscript. We also acknowledge Nicolas Laflorencie, Toni Shiroka, Markos Skoulatos, and Olivier Mentr\'e for discussions and sharing the data prior to publication. A.T. was funded by Alexander von Humboldt Foundation. F.W. acknowledges the assistance of Yurii Skourski during the high-field magnetization measurements and the financial support under the project M.FE.A.CHPHSM of the Max-Planck Society. Part of this work has been supported by EuroMagNET II under the EC contract 228043.

%\bibliography{bicu2po6}

\begin{thebibliography}{68}
\bibitem [{not(){\natexlab{c}}}]{note2}%
  \BibitemOpen
  \bibinfo {note} {The projector on the correlated orbitals was defined such
  that the trace of the occupation number matrices represents the 3$d$ gross
  occupation.}\BibitemShut {Stop}%
\bibitem [{not(){\natexlab{e}}}]{note6}%
  \BibitemOpen
  \bibinfo {note} {LSDA+$U$ does not account for quantum fluctuations which are
  inherent to low-dimensional spin-$\frac12$ systems. Therefore, the value of
  the magnetic moment cannot be compared to the experiment.}\BibitemShut
  {Stop}%
\bibitem [{not(){\natexlab{f}}}]{note3}%
  \BibitemOpen
  \bibinfo {note} {In this analysis, we restrict ourselves to the model
  approach, because the FM contributions to $J_2$ and $J_2'$ are small, see
  Tables~\ref{tbm} and~\ref{exchange}.}\BibitemShut {Stop}%
\bibitem [{not(){\natexlab{g}}}]{note4}%
  \BibitemOpen
  \bibinfo {note} {Note that further rotations lead to a more complex evolution
  of $J_2$ and $J_2'$. This may be caused by the changes in the overall bonding
  situation.}\BibitemShut {Stop}%
\bibitem [{not(){\natexlab{i}}}]{note8}%
  \BibitemOpen
  \bibinfo {note} {The slight curvature of the data above $H_c$ is likely an
  experimental effect due to the low magnetic moment of the
  sample.}\BibitemShut {Stop}%
\bibitem [{not(){\natexlab{j}}}]{note7}%
  \BibitemOpen
  \bibinfo {note} {Here, we imply the simple reflection in the crystallographic
  mirror plane without the time reversal in the spin space.}\BibitemShut
  {Stop}%
\end{thebibliography}
%merlin.mbs 2010-03-15 4.21a (PWD, AO, DPC)
%Control: key (0)
%Control: author (8) initials jnrlst
%Control: editor formatted (1) identically to author
%Control: production of article title (-1) disabled
%Control: page (0) single
%Control: year (1) truncated
%Control: production of eprint (0) enabled
%

\end{document}